# Built Environment and Walking:
# Short vs. Long Walking Trips

**Aryan Hosseinzadeh**[1]

*Department of Civil and Environmental Engineering, University of Louisville, W.S. Speed, Louisville, KY, 40292, USA*

[1] *E-mail: aryan.hosseinzadeh@louisville.edu*

## Abstract

In recent decades, many studies investigated the influencing factors on walking. Although there are lots of finding about these factors, only a few of them conducted to differentiate between short and long walking trips and their associated influencing factors. Current research investigates the impact of the influencing factors on the share of short. To do so, in the first step a proxy between short and long walking trip had been recognized. In this regard, indices mentioned in the literature review are derived from a transportation network database and land use data. According to the results, density in both trip generation sides in short trips is significant. Models are able to describe variation in share of walking up to 0.277 in short return to home walking trips.

## Introduction

Walking as a basic mode of transportation plays a significant role in urban transportation (Reid Ewing & Cervero, 2010; Piatkowski, Krizek, & Handy, 2015; Shaaban, Muley, & Elnashar, 2018; Staats, Diakité, Voûte, & Zlatanova, 2019). Alongside providing access to other motorized modes, walking can be considered as independent travel mode stand alone, especially in shorter trips. Besides transportation, numerous benefits of promoting walking include the environment, economy (Talen & Koschinsky, 2013) and public health (Braun et al., 2016) have been mentioned in the literature.

Current study addresses influencing factors on walking in trip generation zones in short trips by considering the boundary of short and long walking trips as well as different trip purposes. In this study, the effective factors on the share of walking trips are based on 112 Rasht TAZs have been modeled, considering using various indices which have been introduced in the literature review. Due to the possibility of different influences of trip produced and attracted trips, results are reported separately.

The remainder of this paper is organized as follows. The next section provides an overview of indices related to each BE criterion previously proposed and studies which considered the length of walking trips. The third section offers a brief explanation of the method that has been used, an overview of the case study and the descriptive statistics. The results and conclusion are the fourth and final section of the paper.

## Literature review

The first section is a discussion about BE criteria that have been mentioned in the literature.

### Built Environment Criteria

#### Design

Design represents street network characteristics within an area. Quite a few indices have been recommended to encapsulate the effect of transportation network design on walking (Berrigan, Pickle, & Dill, 2010; Dill, 2004; Gori, Nigro, & Petrelli, 2014; Schlossberg, 2006). Based on this method, Hatamzadeh et al. suggested two combinatorial indices that have resulted from principal component analysis (Hatamzadeh, Habibian, & Khodaii, 2016). According to their result, two components were extracted, which are capable of delineating 71.02% of variations of all design indices (Hatamzadeh et al., 2016). The first component consists of percentages of four-way intersections and connected node ratio (node connectivity). The second component includes the ratio of minor streets to major streets and street density (link connectivity) (for more details see (Hatamzadeh et al., 2016). The first part of Table 1 represents the previously employed indices in the literature (Habibian & Hosseinzadeh, 2018).

#### Diversity

Various indices are proposed to measure the diversity of land use. Table 1 shows the definition and determination method of each diversity index in the literature (Habibian & Hosseinzadeh, 2018). Although a large number of studies has implemented the Entropy index, Christian et al. explored the variation of entropy based on considering different types and categorization. They conclude that entropy index could hugely change due to slight modification in land use categorization (Christian et al., 2011).

*Density*

Density as a BE criterion is considered as a ratio of the population of a specific zone to the area of that zone (L. D. Frank et al., 2010).

Table 1 BE indices in the literature review ((Habibian & Hosseinzadeh, 2018))

| | Variable | Description | Impact[2] | References |
|---|---|---|---|---|
| | | Design indices | | |
| 1 | Intersection density | Number of intersections per unit area | + | (Badland et al., 2009; L. D. Frank et al., 2010; L. D. Frank et al., 2005; Holt, Spence, Sehn, & Cutumisu, 2008; Koohsari et al., 2016; McCormack, Cerin, Leslie, Du Toit, & Owen, 2007; McGinn, Evenson, Herring, Huston, & Rodriguez, 2007; Nagel, Carlson, Bosworth, & Michael, 2008; Van Dyck, Deforche, Cardon, & De Bourdeaudhuij, 2009; Wells & Yang, 2008) |
| 2 | Percentage of 4-way intersections | Ratio of 4-way intersections to all intersections × 100 | + | (Dill, 2004) |
| 3 | Cul-de-sac density | Number of cul-de-sacs per unit area | - | (Schlossberg & Brown, 2004) |
| 4 | Pedestrian catchment area | Pedestrian accessible area (PA)/ Ideal pedestrian accessible area (IA) | + | (Chin, Van Niel, Giles-Corti, & Knuiman, 2008; Gori et al., 2014; Porta & Renne, 2005; Schlossberg, 2006; Schlossberg & Brown, 2004) |
| 5 | Modified pedestrian catchment area | Modified pedestrian accessible area (MPA)/Ideal pedestrian accessible area (IA) | + | (Gori et al., 2014) |
| 6 | Impeded pedestrian catchment area | Pedestrian accessible area considering impedances / Ideal pedestrian accessible area (IA) | + | (Schlossberg, 2006) |
| 7 | Ratio of minor streets[3] to major streets[4] | - | + | (Dill, 2004) |
| 8 | Block density | Number of blocks per unit area | + | (Dill, 2004; Hooper, Knuiman, Foster, & Giles-Corti, 2015; Song & Knaap, 2004) |
| 9 | Block length | Average length of blocks in an area | - | (S. Handy, Paterson, & Butler, 2003) |

[2] +, - and * show the positive, negative and contradictory impact of the indices in the previous studies.
[3] Minor street is considered as two-way two-lane urban street that often services to low traffic and low speed.
[4] Major street is considered as two-way four-lane or more urban street (multilane streets) that can serve more vehicles with high average speed.

| | | | | |
|---|---|---|---|---|
| 10 | Street density | Total length of streets per unit area | + | (Dill, 2004) |
| 11 | Connected node ratio (CNR) | Number of intersections divided by the number of intersections plus cul-de-sacs | + | (Berrigan et al., 2010; Dill, 2004; Hooper et al., 2015) |
| 12 | Ratio of link-nodes | Ratio of links to nodes per unit area | + | (Berrigan et al., 2010; Dill, 2004; Zhang & Kukadia, 2005) |
| 13 | Grid pattern | Similarity of a street network to grid network | + | (Southworth & Owens, 1993) |
| 14 | Pedestrian route directness (PRD) | Ratio of route distance to straight-line distance for two selected points | - | (Dill, 2004) |
| 15 | Gamma index | Ratio of number of actual links to the number of all possible links | * | (Berrigan et al., 2010; Dill, 2004; Gori et al., 2014; Schlossberg, 2006; Schlossberg & Brown, 2004) |
| 16 | Alpha index | Ratio of number of actual loops to the number of all possible loops | * | (Berrigan et al., 2010; Dill, 2004; Gori et al., 2014; Schlossberg, 2006; Schlossberg & Brown, 2004) |
| 17 | Node connectivity | 0.817 (Percentage of four-way intersection) + 0.817 (The ratio of intersection to nodes) | * | (Hatamzadeh et al. 2017) |
| 18 | Link connectivity | 0.862 (The ratio of minor roads to major roads) + 0.762 (street density) | + | (Hatamzadeh et al. 2017) |

| | Diversity indices | | | |
|---|---|---|---|---|
| 1 | Entropy | $-\dfrac{\sum_{i=1}^{n} p_i \log p_i}{\log n}$ <br> $P_i$: Percentage of land use i (area-based) <br> n: Total number of land uses | * | (Cervero & Kockelman, 1997; L. D. Frank et al., 2010; L. D. Frank et al., 2005; Taleai & Amiri, 2017) |
| 2 | Herfindal-Hershman index (HHI) | $p_1^2 + p_2^2 + \cdots + p_n^2$ <br> $P_i$: Percentage of land use type i <br> n: Total number of land uses | - | (Eriksson, Arvidsson, Gebel, Ohlsson, & Sundquist, 2012) |
| 3 | MXI | $|P - 50|$ <br> P: Percentage of residential land use of a specific area | - | (Van den Hoek, 2008) |
| 4 | Job-population balance | $1 - \left| \dfrac{Job - 0.2 \times Pop}{Job + 0.2 \times Pop} \right|$ <br> job: Number of jobs within a specific area <br> pop: Number of residents within a specific area | + | (R. Ewing et al., 2014) |
| 5 | Dissimilarity index | $\dfrac{X_i}{8}$ | + | (Cervero & Kockelman, 1997) |

$X_i$: Number of dissimilar land uses adjacent to a considered land use

*Distance to Transit*

Previous studies use various indices to capture distance to transit criteria.

*Destination Accessibility*

Destination accessibility defines as ease of accessing different destination locations.

## Case study

### Area of Study

The city of Rasht (population about 640 thousand in 2007) is the largest city on Iran's Caspian Sea coast (the north of Iran). The urban area in Rasht includes 112 traffic analysis zones (TAZs), which are shown in figure 1. Unplanned settlements with disordered pathways, dense residential and weak infrastructure form a significant part of the spatial structure in the city (Hatamzadeh, Habibian, & Khodaii, 2017).

### Data Description

In this study, the information of Rasht Household Travel Survey (RHTS) in 2007 is used[5]. As a part of the study, a questionnaire was designed and distributed among more than 5000 households who reside in 112 TAZs. The data consists of more than 5000 household and 17000 individuals and 30000 trips (5501 work trips, 4896 educational trips, 2737 shopping trips and 15355 return to home trips). The data description is mentioned in Table 2.

---

[5] It is worth mentioning that as RHTS has not been updated since 2007, no more recent data has been available

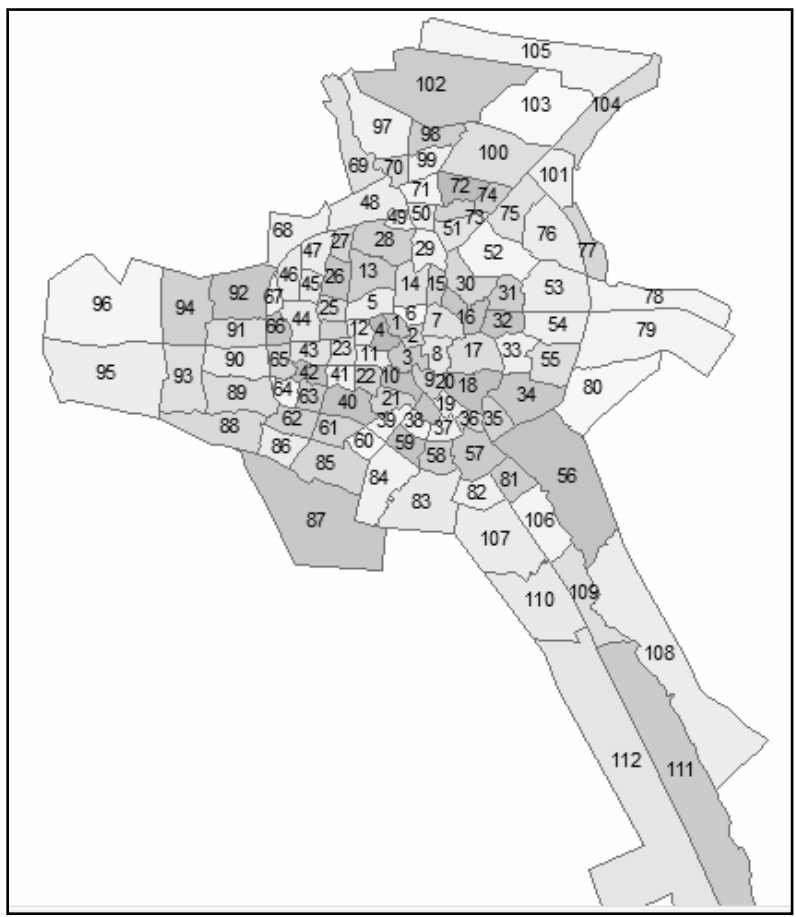

Figure 1. Traffic analysis zones in Rasht

Table 2. Zone-based descriptive statistics

|  |  | Average | Standard deviation | Min | Max |  |
|---|---|---|---|---|---|---|
| Part 1- Socio-economic variables | | | | | | |
| 1-1 | Age average | 30.01 | 2.83 | 22.33 | 40.25 | |
| 1-2 | Household size | 3.51 | 0.19 | 3 | 4.33 | |
| 1-3 | Average bike ownership | 0.66 | 0.12 | 0.375 | 1.008 | |
| 1-4 | Average motor ownership | 0.12 | 0.07 | 0 | 0.43 | |
| 1-5 | Average car ownership | 0.57 | 0.24 | 0.26 | 0.95 | |
| Part 2- Connectivity indices | | | | | | |
| 2-1 | Intersection density | 244.5 | 134.32 | 1.24 | 656.29 | $1/km^2$ |
| 2-2 | Percentage of 4-way intersections | 14.1 | 6.44 | 0 | 38.9 | |
| 2-3 | Cul-de-sac density | 146.11 | 91.01 | 0 | 407.8 | $1/km^2$ |
| 2-4 | Number of cul-de-sac | 75 | 62.6 | 0 | 363 | |
| 2-5 | Number of 3-way | 104 | 80.6 | 3 | 449 | |
| 2-6 | Number of 4-way | 15 | 11.96 | 0 | 69 | |

| 2-7 | Ratio of minor streets to major streets | 11.57 | 25.63 | 0 | 187.56 | |
| 2-8 | Street density | 0.017 | 0.0084 | 0.004 | 0.035 | m/km$^2$ |
| 2-9 | 3-way intersection density | 210.88 | 116.92 | 1.24 | 535 | 1/km$^2$ |
| 2-10 | 4-way intersection density | 33.62 | 23.98 | 0 | 121.2 | 1/km$^2$ |
| 2-11 | Connected node ratio | 0.62 | 0.1 | 0.4 | 1 | |
| 2-12 | Ratio of links to nodes | 1.86 | 0.2 | 1.55 | 2.25 | 1/m |
| 2-13 | Gamma index | 0.39 | 0.05 | 0.33 | 0.63 | |
| 2-14 | Alpha index | 0.09 | 0.059 | 0.01 | 0.36 | |
| 2-15 | Percentage of 3-way intersections | 85.9 | 10.32 | 61.09 | 100 | |
| 2-16 | Number of major 3-way intersections | 7.61 | 7.48 | 0 | 49 | |
| 2-17 | Number of major 4-way intersections | 1.34 | 1.67 | 0 | 11 | |
| 2-18 | Ratio of cul-de-sac to nodes | 37.01 | 9.22 | 0 | 60 | |
| 2-19 | Major street density | 3653.4 | 2967.5 | 0 | 16149.4 | m/km$^2$ |
| 2-20 | Minor street density | 21648 | 8836.3 | 610 | 36898 | m/km$^2$ |
| 2-21 | Average link length | 54.37 | 24.95 | 27.2 | 227.46 | m |
| **Part3-Diversity indices** | | | | | | |
| 3-1 | Entropy index | 0.33 | 0.19 | 0 | 0.83 | |
| 3-2 | HHI | 0.72 | 0.18 | 0.29 | 1 | |
| 3-3 | MXI | 35.8 | 11.3 | 1.64 | 50 | |
| 3-4 | Job-pop balance | 0.56 | 0.29 | 0 | 1 | |
| **Part4-Density index** | | | | | | |
| 4-1 | Population density | 10100 | 6600 | 0 | 28700 | 1/km$^2$ |
| **Part5-Destination accessibility indices** | | | | | | |
| 5-1 | Areal distance to CBD | 2629 | 1712 | 0 | 10626 | m |
| 5-2 | Network distance to CBD | 3334 | 2318 | 0 | 14782 | m |

## Methodology

In this study, linear regression analysis is used to modeling the influencing factors on the share of short walking trips. Share of short walking trips in a zone is considered as a dependent variable and socio-demographic characteristics and built environment criteria are used as independent variables. Thus, the model can be represented in equation 1:

$$Y_i = \beta_1 + \beta_2 x_{2i} + \beta_3 x_{3i} + \ldots + u_i \qquad (1)$$

Given a data set of n TAZs, a linear regression model assumes that the relationship between the share of walking $Y_i$ and each of the independent variables is linear (equation 1). This relationship modeled through an error term $u_i$, an unobserved random variable that adds noise to equation 1.

In developing equation 1, it has assumed equation2, no serial correlation (equation 3), homoscedasticity of the erorr term (equation 4), zero covariance between $u_i$ and each $x_i$ variables (equation 5), no specification bias and no exact collinearity between the x variables.

$$E(\,u_i \mid X_{2i}, X_{3i}) = 0 \text{ for each } i \qquad (2)$$

$$Cov(u_i,\, u_j) = 0 \qquad (3)$$

$$Var(u_i) = \sigma^2 \qquad (4)$$

$$Cov\,(u_i, X_{2i}) = Cov\,(u_i, X_{3i}) = 0 \qquad (5)$$

The goodness of fit and adjusted goodness of fit for the model resulted from equation 6 and equation 7.

$$R^2 = \frac{ESS}{TSS} = \frac{\widehat{\beta_2}\sum y_i x_{2i} + \widehat{\beta_3}\sum y_i x_{3i} +}{\sum y_i^2} \qquad (6)$$

$$\overline{R^2} = 1 - \frac{\sum \widehat{u}_i^2 / (n-k)}{\sum y_i^2 / (n-1)} \qquad (7)$$

## Result and Discussion

### *Walking Trip Production*

In this part, walking trip production models are represented and discussed.

Table 4. Walking trip production models across various trip purposes[6]

|  | Short trips |
|---|---|
|  | Work trips |
| Constant | 0.023*** |
| Population Density | 0.00042*** |
| Link Connectivity | - |
| $R^2$ | 0.123 |
| Adj-$R^2$ | 0.115 |
|  | Educational trips |

---

[6] ***, ** and * means 99%, 95% and 90% level of significance, respectively.

| | |
|---|---|
| Constant | 0.196** |
| Population Density | 0.00048** |
| Car ownership | - |
| Link Connectivity | - |
| Average link length | -0.0022* |
| $R^2$ | 0.111 |
| Adj-$R^2$ | 0.094 |
| Shopping trips | |
| Constant | 0.102*** |
| Population Density | 0.0009*** |
| Car ownership | - |
| Job-pop balance | - |
| Link Connectivity | 0.068*** |
| $R^2$ | 0.288 |
| Adj-$R^2$ | 0.274 |
| Return to home trips | |
| Constant | -0.017*** |
| Population density | 0.001*** |
| Job-pop balance | 0.151*** |
| $R^2$ | 0.29 |
| Adj-$R^2$ | 0.277 |

According to Table 4, population density in all short walking trips production is significant. In educational and shopping trips (which usually their production zone is in individuals' home), car ownership found significant. It means individuals prefer to use their car in longer trips which is intuitive. This would be more influential in shopping trips since the long length of trip and carry stuff would be hard for a walker. Connectivity measures in short educational and shopping trips found significant.

As it shows in the models, BE criteria are more successful in describing walking trip production in shopping and return to home short walking trips. The reason could be individuals more free time on these trips versus work and educational trips which shorter distances motivate them to walk.

*Walking Trip Attraction*

In this part, walking trip attraction models represented and discussed.

**Table 5. Walking trip attraction models across various trip purposes**

|  | Short trips |
| --- | --- |
| Constant | 0.16*** |
| Population density | 0.0004*** |
| Job-pop balance | 0.086** |
| $R^2$ | 0.175 |
| Adj-$R^2$ | 0.16 |
| Constant | 0.006*** |
| Population density | 0.001*** |
| Job-pop balance | 0.123** |
| Average link length | - |
| $R^2$ | 0.112 |
| Adj-$R^2$ | 0.096 |
| Constant | 0.045 |
| Population density | 0.002*** |
| Job-pop balance | 0.209* |
| $R^2$ | 0.164 |
| Adj-R$^2$ | 0.147 |
| Constant | 0.113*** |
| population density | 0.001*** |
| Job-pop balance | - |
| Average link length | -0.001** |
| R$^2$ | 0.299 |
| Adj-R$^2$ | 0.286 |

Job-population balance, as a proxy of diversity, shows if there is a balance between job opportunities and residents in a zone. The positive sign of the index in the models means this balance has a positive impact on increasing walking share in a zone.

## Conclusion

Population density is significant in all eight produced and attracted models of short walking trips. This finding perfectly highlights the importance of density in short walking trips. Beside the previous studies, mostly consider the population density just in residual place of the respondents (L. D. Frank et al., 2010; L. D. Frank et al., 2005; Talen & Koschinsky, 2013).

Shopping and return to home short walking trips resulted in the higher goodness of fit comparing work and educational trips. The reason could be the individuals free time in shopping and return to home trips. Work and educational trips may limit individuals to be a specific place in a fixed time.